\documentclass[lettersize,journal]{IEEEtran}
\usepackage[T1]{fontenc}
\usepackage[utf8]{inputenc}
\usepackage{amsmath,amsthm,mathtools,amssymb,amsfonts}
\usepackage{dsfont}
\usepackage{graphicx}
\usepackage{units}
\usepackage{booktabs}          % modern tables
\usepackage{array}

\usepackage{algorithm}
\usepackage[noend]{algpseudocode}

\usepackage{enumitem}
\usepackage{pifont}

\newcommand{\matr}[1]{{\boldsymbol{#1}}}
\renewcommand{\vec}[1]{{\boldsymbol{#1}}}

\newcommand{\nn}{\mathtt{N}}

\newcommand{\imu}{{\rm j}}
\newcommand{\eye}{\matr I}%{\mathds{1}}

\usepackage{url}

\usepackage[breaklinks,hidelinks]{hyperref}

\usepackage{orcidlink}

\begin{document}

\title{Low-rank corrections to a shared reference Jacobian for GPU-parallel AC $N\!-\!1$ contingency analysis
\thanks{M.H., D.W. and L.S. acknowledge financial support from the Elia Group via the Elia Group Open Research Challenge 2025. The authors gratefully acknowledge computing time on the supercomputer JURECA~\cite{JURECA} at Forschungszentrum Jülich under grant no. gpu-to-power.}
}  

\author{
Marc Hunkem\"oller\,\orcidlink{0009-0004-6494-1456},
Nico Westerbeck\,\orcidlink{0009-0004-1761-4305},
Joost van Dijk\,\orcidlink{0009-0003-1616-3621},
Lars Schewe\,\orcidlink{0000-0002-3778-262X}
and~Dirk Witthaut\,\orcidlink{0000-0002-3623-5341}%
\thanks{M. Hunkem\"oller and D. Witthaut are with the Institute of Climate and Energy Systems (ICE-1), Forschungszentrum J\"ulich, 52428 J\"ulich, Germany, and also with the Institute for Theoretical Physics, University of Cologne, 50937 K\"oln, Germany (e-mail: m.hunkemoeller@fz-juelich.de; d.witthaut@fz-juelich.de).}%
\thanks{L. Schewe is with the School of Mathematics and Maxwell Institute for Mathematical Sciences, University of Edinburgh, Edinburgh EH9 3FD, U.K. (e-mail: lars.schewe@ed.ac.uk).}%
\thanks{J. van Dijk is with TenneT TSO BV, 6812 AR Arnhem, The Netherlands (e-mail: joost.van.dijk@tennet.eu).}%
\thanks{N. Westerbeck is with Elia Group, 1000 Brussels, Belgium (e-mail: nico.westerbeck@50hertz.com).}%
}

\maketitle
 
\begin{abstract}
Contingency analysis repeatedly solves the power-flow equations under many single-element outages, an increasingly demanding workload as operating margins tighten and renewable variability grows. Graphics processing units (GPUs) are attractive for this task, but a change in topology alters the sparsity pattern of the Jacobian, preventing the uniform batched factorization on which efficient GPU solvers rely. This paper presents a GPU-parallel Newton--Raphson solver for $N\!-\!1$ contingency analysis that separates the change in operating state from the change in topology. A single reference Jacobian is inverted once and reused across all contingencies. Each topology change is expressed as a low-rank Woodbury correction, while the residual state-dependent perturbation is handled by an inner iterative solve, a truncated Neumann series or the biconjugate gradient stabilized method. We analyze the spectral radius and condition number of the resulting iteration matrices and show that the reference Jacobian is an effective preconditioner. On transmission-scale benchmarks, the method attains per-topology solve times competitive with an optimized 80-core central processing unit baseline, with the low-rank correction accounting for approximately 1--2\% of the workload. This makes GPU-parallel $N\!-\!1$ contingency analysis feasible at transmission scale without per-topology refactorization.

\end{abstract}


\section{Introduction}
\label{sec:introduction}

\IEEEPARstart{F}{ast} and reliable power-flow (PF) computation is essential for power-system operation and planning. It is used in state estimation, contingency screening and security assessment and appears as the repeated inner solve in optimization tasks such as optimal power flow and topology optimization~\cite{Grai94, Wood14, kundur2022power}. These demands intensify with the energy transition: growing shares of volatile, converter-based generation, tighter operating margins from demand growth outpacing transmission expansion and more frequent corrective actions require evaluating large numbers of PF scenarios under time constraints~\cite{Simp16, witthaut2022collective}.

For accurate AC analysis, Newton--Raphson (NR) methods combined with sparse linear algebra remain the standard, owing to their robustness and fast local convergence~\cite{Grai94, Wood14, kundur2022power}. In many operational settings, however, the computational bottleneck is not a single NR solve but
the repeated solution of many closely related PF problems, most prominently $N\!-\!1$ contingency analysis, in which every single-element outage of a system with $N$ components must be assessed. Because the number of contingencies grows with system size, this workload, combined with the high similarity between contingency cases (only one element changes), is a natural target for parallel hardware such as graphics processing units (GPUs). However, the obstacle is structural. Varying nodal injections preserves the Jacobian’s sparsity pattern, so a single symbolic factorization can be reused. A topology change, however, alters this pattern. The resulting fill-in and irregular memory access make sparse direct solvers a poor fit for the single-instruction, multiple-data (SIMD) parallelism of GPUs, limiting their application across topologies.

In this work we develop an AC PF solver that resolves this obstacle by separating the Jacobian's dependence on the system state from its dependence on the topology. Each local topology modification is represented as a structured low-rank correction to a fixed-sparsity reference Jacobian, whose inverse is computed once and reused across all contingencies. The low-rank correction is applied at solve time through the Woodbury matrix identity, thus avoiding a refactorization, while the remaining state-dependent perturbation is resolved by an iterative inner solve, either a truncated Neumann series or biconjugate gradient stabilized (\mbox{Bi-CGSTAB}) method. Then, every contingency reduces to batched dense and sparse matrix products with shared sparsity pattern. Thus, the entire $N\!-\!1$ study maps efficiently onto GPU hardware while retaining the accuracy and robustness of a full AC NR solve.

The remainder of this paper is organized as follows. Section~\ref{sec:litreview} reviews related work on linearized screening, low-rank modification techniques, GPU acceleration and iterative solvers for PF. Section~\ref{sec:background} recalls the AC PF equations, the low-rank structure of the Jacobian under topology changes and the numerical tools we build on. Sections~\ref{sec:jacobian-decompositions} and~\ref{sec:algorithm} present the Jacobian decompositions and the resulting GPU-parallel algorithms. Section~\ref{sec:spectral} analyzes the spectral properties that govern their convergence and selects the best variant, which Section~\ref{sec:performance} benchmarks against multi-core central processing unit (CPU) baselines on transmission-scale networks. Section~\ref{sec:conclusion} concludes. 

\section{Literature review}
\label{sec:litreview}

\subsection{Linearized and screening methods}
For computational efficiency, linearized models such as DC PF and its refinements are widely used~\cite{Stot09, purchala2005usefulness}, with extensions incorporating losses, reactive power and voltage magnitudes~\cite{Coff12, Li2018, Yang2019}. Distribution-factor methods in particular enable extremely fast contingency and topology screening~\cite{guler2007generalized, ronellenfitsch2017dual, van2024unified}, because a topology change enters as a low-rank update to a fixed linear operator and recent formulations extend these factors to capture voltage sensitivity~\cite{titz25}. However, the linear approximation can break down in stressed operating regimes where the nonlinear AC effects in the contingency assessment are no longer negligible. 

\subsection{Low-rank modifications and factorization updates}
The observation that a single-element outage is a low-rank change to the system matrix has a long history in power engineering, dating back to compensation-based techniques for handling network modifications without refactorizing~\cite{Alsac83}. In numerical linear algebra, efficient algorithms for low-rank updates of sparse LU factorizations are well established~\cite{stange2006efficient, saunders2009lusol}. Direct application of such factor-update schemes on GPUs is nontrivial, because the update algorithms pivot across rows and columns, requiring data movement in GPU memory that undermines throughput. And, for contingency analysis, the full updated factorization is not even needed, only the solution of the corresponding linear system. This motivates applying the low-rank correction at solve time rather than updating the factors, for which the Woodbury matrix identity provides the natural tool~\cite{hager1989updating, higham2002accuracy}.
In the DC approximation, low-rank updates of the power transfer distribution factors can represent line outages and substation reconfigurations without refactorization, achieving extreme throughput on GPUs~\cite{westerbeck2025accelerated}. This demonstrates the potential of the low-rank viewpoint for AC PF with topology changes.

\subsection{GPU acceleration for power flow}

GPUs have been widely used to accelerate the many PF solves required when numerous operating scenarios share a single network topology. The typical case is probabilistic power flow, where a Monte-Carlo ensemble of nodal injections, e.g., to reflect the uncertainty of renewable generation and load, is evaluated on a fixed grid~\cite{zhou2018online, wang2021fast}. Time-series simulation and scenario-based planning studies fall into the same category. In all of these studies, only the injections vary while the topology and hence the Jacobian sparsity pattern, stays fixed. The workload therefore reduces to a batch of structurally identical linear solves, which maps naturally onto the GPU through batched sparse factorization or batched matrix operations~\cite{su2020full, ahmadi2022fast}.

The situation is fundamentally different for $N\!-\!1$ contingency analysis and topology optimization, where each scenario removes or modifies a network element. Such a change alters the sparsity pattern of the Jacobian, so the fixed-pattern batching of the multi-scenario case no longer applies directly and sparse direct solvers become sensitive to the resulting fill-in and irregular memory access. This is addressed by regenerating and refactorizing many Jacobians in parallel on the GPU for static security analysis~\cite{zhou2017gpuaccelerated}, which recovers parallelism but repeats the factorization for every topology. Compensation-type methods treat a branch outage as a low-rank update to the constant fast-decoupled matrices, reusing a single factorization across all contingencies on parallel hardware~\cite{huang2018realtime}. We instead separate the state and topology dependence of the full AC Jacobian and apply the low-rank correction at solve time.%, so that every contingency reduces to batched dense and sparse matrix products on a fixed sparsity pattern.

\subsection{Iterative solvers for power flow}
Iterative solvers are a long-standing alternative to direct factorization for the linear systems arising in each NR PF solve. Krylov-subspace methods were introduced to PF computation as a scalable alternative to LU factorization~\cite{Semlyen1996fundamentals}. Subsequent work combined them with NR in inexact-Newton and recycling schemes, comparing inner solvers such as the generalized minimal residual (GMRES), \mbox{Bi-CGSTAB} and Quasi-Minimal Residual (QMR) methods and reusing preconditioner or subspace information across NR iterations~\cite{yetkin2023recycling}. Fast-decoupled and conjugate-gradient formulations that retain a largely fixed sparsity pattern have likewise been proposed, avoiding the refactorization that a topology-induced change in matrix structure would otherwise require~\cite{meng2023fastdecoupled, fu2020conjugategradiant}. The appeal of these methods on parallel hardware is that the dominant operation is the matrix-vector product, which parallelizes far more readily than the inherently sequential triangular solves of a direct method.

\subsection{Fair benchmarking of GPU speedups}
Reported GPU speedups for PF computation span nearly two orders of magnitude, largely determined by the choice of CPU baseline. High speedups are often reported relative to general-purpose open-source tools~\cite{wang2021fast}, while comparisons against optimized sparse solvers can yield more modest figures~\cite{dorto2021comparing}. This mirrors the more general finding that, once both sides are properly optimized and parallelized, the GPU advantage shrinks substantially relative to the orders of magnitude frequently claimed~\cite{lee2010debunking}. We therefore benchmark against multi-core CPU baselines rather than a single-threaded reference seeking a CPU baseline of comparable cost and capability rather than an artificially weak one.

\section{Background}
\label{sec:background}

\subsection{Power-flow equations and Newton--Raphson method}
\label{sec:background-pf}

The AC PF equations describe the steady state of a power grid. They relate the real and reactive power injections $P_i$ and $Q_i$ at node $i$ to the voltage magnitudes and phase angles at all nodes. We work with logarithmic voltage variables $u_i = \log v_i$, which leads to a compact Jacobian structure~\cite{milano2022complex}. The phase angles are denoted as $\theta_i$. With the shorthand $\theta_{ij} = \theta_i - \theta_j$ and the nodal admittance matrix entries $G_{ij} + \imu B_{ij}$, the PF equations read
\begin{align}
    P_i &= \sum_{j} e^{u_i} e^{u_j}
       \left( G_{ij} \cos\theta_{ij} + B_{ij} \sin\theta_{ij} \right)
       =: p_i(\vec{x}),
       \label{eq:loadflow_P}\\
    Q_i &= \sum_{j} e^{u_i} e^{u_j}
       \left( G_{ij} \sin\theta_{ij} - B_{ij} \cos\theta_{ij} \right)
       =: q_i(\vec{x}).
       \label{eq:loadflow_Q}
\end{align}
We distinguish $n$ PQ nodes (loads) with fixed $(P_i, Q_i)$, $m$ PV nodes (generators) with fixed $P_i$ and voltage magnitude $u_i$ and one slack node with fixed voltage magnitude and $\theta = 0$. Indexing PQ nodes by $i = 1,\ldots,n$, PV nodes by $i = n+1,\ldots,n+m$ and the slack node by $n+m+1$, the network has $\nn = n+m+1$ nodes and $\ell$ branches.

Collecting the free state variables as $\vec{x} = (\theta_1, \ldots, \theta_{n+m}, u_1, \ldots, u_n)^\top \in \mathbb{R}^{N_m}$ with $N_m = 2n+m$ and the corresponding power mismatches as $\vec{f}(\vec{x}) = (p_1(\vec{x})\!-\!P_1, \ldots, q_1(\vec{x})\!-\!Q_1, \ldots)^\top$, the PF problem takes the compact form $\vec{f}(\vec{x}) = \vec{0}$. The NR method solves this system iteratively. Expanding $\vec{f}$ to first order about the current iterate $\vec{x}^{(\sigma)}$ yields the linear update equations
\begin{align}
    \matr{J}(\vec{x}^{(\sigma)})\, \Delta\vec{x}^{(\sigma)}
    &= -\vec{f}(\vec{x}^{(\sigma)}),
    \label{eq:nr-step}\\
    \vec{x}^{(\sigma+1)} &= \vec{x}^{(\sigma)} + \Delta\vec{x}^{(\sigma)},
\end{align}
where $\sigma$ is the iteration index and $\matr{J} \in \mathbb{R}^{N_m \times N_m}$ is the Jacobian.

To expose the structure that the low-rank correction exploits, we fix an orientation for each of the $\ell$ branches and define the signed node-edge incidence matrix $\matr{E} \in \mathbb{R}^{\nn \times \ell}$, with {$E_{i,a} = +1$ ($-1$)} if branch $a$ originates (terminates) at node $i$ and $0$ otherwise, together with its unsigned counterpart ${D_{i,a} = |E_{i,a}|}$. Throughout, a tilde ($\tilde{\cdot}$) denotes omission of the slack entry and a hat ($\hat{\cdot}$) omission of the slack and PV entries, applied consistently to matrices (rows) and vectors (components). In terms of these operators we define the per-branch quantities
\begin{align*}
    \vec{c} = \cos(\matr{E}^\top \vec{\theta}) \odot \exp(\matr{D}^\top \vec{u}),
    \,\,\,
    \vec{s} = \sin(\matr{E}^\top \vec{\theta}) \odot \exp(\matr{D}^\top \vec{u}),
\end{align*}
where $\odot$ denotes the element-wise multiplication. With ${\vec{b}, \vec{g} \in \mathbb{R}^{\ell}}$ collecting the series susceptance and conductance of each branch, ordered consistently with the edge list, the four Jacobian blocks take the compact form
\begin{align}
    \matr{J}_{p\theta} &=
        \matr{\tilde{E}} [\vec{b} \odot \vec{c}] \matr{\tilde{E}}^\top
        - \matr{\tilde{D}} [\vec{g} \odot \vec{s}] \matr{\tilde{E}}^\top,
        \nonumber\\
    \matr{J}_{q\theta} &=
        \matr{\hat{D}} [\vec{b} \odot \vec{s}] \matr{\tilde{E}}^\top
        + \matr{\hat{E}} [\vec{g} \odot \vec{c}] \matr{\tilde{E}}^\top,
        \nonumber\\
    \matr{J}_{pu} &=
        \matr{\tilde{E}} [\vec{b} \odot \vec{s}] \matr{\hat{D}}^\top
        + \matr{\tilde{D}} [\vec{g} \odot \vec{c}] \matr{\hat{D}}^\top
        + [2\vec{\tilde{\gamma}} \odot \exp(2\vec{\tilde{u}})],
        \nonumber\\
    \matr{J}_{qu} &=
        \matr{\hat{E}} [\vec{g} \odot \vec{s}] \matr{\hat{D}}^\top
        - \matr{\hat{D}} [\vec{b} \odot \vec{c}] \matr{\hat{D}}^\top
        - [2\vec{\hat{\beta}} \odot \exp(2\vec{\hat{u}})],
    \label{eq:jacobian-blocks}
\end{align}
with the full Jacobian being $\matr{J} = \left(\begin{smallmatrix} \matr{J}_{p\theta} & \matr{J}_{pu}\\ \matr{J}_{q\theta} & \matr{J}_{qu}\end{smallmatrix}\right)$. Here $[\cdot]$ denotes the diagonal matrix formed from a vector and $\vec \beta$, $\vec \gamma$ collect the diagonal susceptance and conductance entries of the bus admittance matrix.

\subsection{Low-rank Jacobian structure under topology changes}
\label{sec:background-lowrank}

We want to solve~\eqref{eq:nr-step} in parallel for a set of grid topologies $\mathcal{T}$, for instance all single-line outage scenarios in an {$N\!-\!1$} security assessment. The central obstacle is that topology changes alter the sparsity pattern of $\matr{J}$, which prevents efficient batched parallel factorization with a shared symbolic structure.

The key observation is that the change is of low rank. We fix a reference topology $r$ and consider a contingency topology $t \in \mathcal{T}$ that differs from $r$ by the outage of a single branch $a = (i, j)$. The series susceptance and conductance of branch $a$ change by $\Delta b_a$ and $\Delta g_a$, respectively. Similarly, the diagonal susceptance and conductance entries $\beta$, $\gamma$ at the terminal nodes $i,j$ change by $\Delta\beta_i,\Delta\beta_j$ and $\Delta\gamma_i,\Delta\gamma_j$. The resulting change in the Jacobian has a rank-four structure,
\begin{align}
    \matr{J}_t(\vec{x}) - \matr{J}_r(\vec{x})
    = \matr{R}_t \, \matr{K}_t(\vec{x}) \, \matr{R}_t^\top,
    \label{eq:lowrank}
\end{align}
where $\matr{R}_t \in \mathbb{R}^{(2n+m) \times 4}$ is a sparse matrix % encoding the positions of the terminal nodes of branch $a$, independent of the state $\vec{x}$, is 
defined with the standard basis vectors ${\vec{r}_a = \vec{e}_i}$ and $\vec{s}_a = \vec{e}_j$ as,
\begin{align*}
    \matr{R}_t = 
    \begin{pmatrix}
    \vec{r}_a(1\!:\!n\!+\!m) & \vec{s}_a(1\!:\!n\!+\!m) & \vec{0} & \vec{0}\\
    \vec{0} & \vec{0} & \vec{r}_a(1\!:\!n) & \vec{s}_a(1\!:\!n)
    \end{pmatrix},
    % \label{eq:Rmatrix}
\end{align*}
with $\vec{r}_a(1\!:\!k)$ denoting the projection onto the first $k$ components. And $\matr{K}_t(\vec{x}) \in \mathbb{R}^{4\times 4}$ contains the admittance changes and depends on the state, as defined in \eqref{eq:K_definition_wide}.
\begin{figure*}[!tb]
\normalsize
\begin{align}
\label{eq:K_definition_wide}
    \matr K = 
    \begin{pmatrix}
    + \Delta b_a c_a - \Delta g_a s_a  &
    - \Delta b_a c_a + \Delta g_a s_a  &
    + \Delta b_a s_a + \Delta g_a c_a
    + 2 \Delta \gamma_i e^{2u_i} &
    + \Delta b_a s_a + \Delta g_a c_a \\
    - \Delta b_a c_a - \Delta g_a s_a  & 
    + \Delta b_a c_a + \Delta g_a s_a  &
    - \Delta b_a s_a + \Delta g_a c_a &
    - \Delta b_a s_a + \Delta g_a c_a 
    + 2 \Delta \gamma_j e^{2u_j} \\
    + \Delta b_a s_a + \Delta g_a c_a  &
    - \Delta b_a s_a - \Delta g_a c_a  &
    + \Delta g_a s_a - \Delta b_a c_a  
        - 2 \Delta \beta_i e^{2u_i} &
    + \Delta g_a s_a - \Delta b_a c_a   \\
    + \Delta b_a s_a - \Delta g_a c_a  &
    - \Delta b_a s_a + \Delta g_a c_a  &
    - \Delta g_a s_a - \Delta b_a c_a  &
    - \Delta g_a s_a - \Delta b_a c_a  
    - 2 \Delta \beta_j e^{2u_j}   \\
    \end{pmatrix}
\end{align}
\hrulefill
\vspace*{4pt}
\end{figure*}
For more details on the low-rank-correction structure of the Jacobian matrix see~\cite{osmses2026}.

\subsection{Mathematical tools}
\label{sec:background-tools}

Three tools are central to the algorithm developed in this paper: the Woodbury matrix identity and iterative solvers based on the Neumann series and {Bi-CGSTAB}.

\subsubsection{Woodbury matrix identity}

For conformable matrices $\matr{M}$, $\matr{U}$, $\matr{C}$ and
$\matr{V}$, the inverse of a low-rank-corrected matrix satisfies
\begin{multline}
    (\matr{M} + \matr{U}\matr{C}\matr{V})^{-1} \\
    = \matr{M}^{-1} - \matr{M}^{-1}\matr{U} \left(\matr{C}^{-1} + \matr{V}\matr{M}^{-1}\matr{U}\right)^{-1} \matr{V}\matr{M}^{-1}\\
    = \matr{M}^{-1} - \matr{M}^{-1}\matr{U} \left(\eye + \matr C \matr{V}\matr{M}^{-1}\matr{U}\right)^{-1} \matr C \matr{V}\matr{M}^{-1},
    \label{eq:woodbury}
\end{multline}
provided the relevant inverses exist~\cite{hager1989updating}. Here, $\eye$ denotes the identity matrix. This avoids inverting the full corrected matrix: if $\matr M^{-1}$ is known, the only remaining inversion is of the small matrix $\matr{C}^{-1} + \matr{V}\matr M^{-1}\matr{U}$, whose size equals the rank of the correction and is therefore cheap. In the present context, $\matr M = \matr{J}_r(\vec{x}_r)$ and the correction has rank four, so the inversion reduces to a $4\times 4$ system.

We apply $\matr{J}_r(\vec{x}_r)^{-1}$ by storing the dense inverse explicitly, computed once at the start of the algorithm. This one-time inversion carries the dominant computational cost. The subsequent application to any right-hand side is a dense matrix--vector (or matrix--matrix) product in $\mathcal{O}(N_m^2)$ operations per column, which maps efficiently onto the GPU.

This choice runs counter to common practice on CPUs, where an explicit dense inverse is avoided in favor of a sparse LU factorization and triangular solves. On the GPU, however, the regular, branch-free dense products underlying the explicit dense inverse achieve far higher throughput than the irregular, sequential memory access of sparse triangular solves, particularly when many right-hand sides are processed simultaneously. The dense inverse also carries a cost: it is no longer sparse, with a memory footprint of $\mathcal{O}(N_m^2)$, which remains modest for the systems considered.

Because the inverse will serve only as a preconditioner, it does not need to be stored in full double precision. Single precision halves its memory footprint and substantially reduces the cost of the dense products. In addition, an explicit inverse is less numerically stable than a factorization~\cite{higham2002accuracy}, more so at reduced precision and as the condition number grows with system size, thus some accuracy is lost in the preconditioner application. As it only preconditions an inner iterative solve rather than providing the final solution, this loss is tolerable.

\subsubsection{Neumann series}
The inverse of a matrix $(\eye - \matr{A})$ can be expanded as a power series,
\begin{equation}
    (\eye - \matr{A})^{-1} = \sum_{k=0}^{\infty} \matr{A}^k,
    \label{eq:neumann}
\end{equation}
provided the spectral radius satisfies $\rho(\matr{A}) < 1$~\cite{horn2012matrix}. The series converges geometrically with asymptotic factor $\rho(\matr A)$, so it converges rapidly when $\rho \ll 1$. Thus it can be truncated after $N_k$ terms with a truncation error that decays asymptotically as $\rho(\matr{A})^{N_k}$ \cite{varga2000matrix, saad2003iterative}.

\subsubsection{Biconjugate gradient stabilized method}

The biconjugate gradient stabilized (\mbox{Bi-CGSTAB}) method~\cite{vandervorst1992BiCGSTAB} solves a linear system $\matr B \vec y = \vec b$ iteratively and, as a Krylov subspace method, applies to any nonsingular $\matr B$ without requiring a spectral radius condition. We use Bi-CGSTAB rather than the conjugate gradient (CG) method because the PF Jacobian is nonsymmetric and CG is restricted to symmetric positive-definite systems. For a symmetric positive-definite system, the convergence of CG is rigorously bounded in terms of the condition number $\kappa$ alone, decreasing by a factor ${\left(\sqrt{\kappa}-1\right)/\left(\sqrt{\kappa}+1\right)}$ per iteration~\cite{saad2003iterative}. For a nonsymmetric operator no such clear bound exists: the convergence of Bi-CGSTAB depends on the full spectrum~\cite{greenbaum1996any,trefethen2005spectra} and the condition number can only give a coarse estimate.

The efficiency of \mbox{Bi-CGSTAB} depends strongly on preconditioning~\cite{saad2003iterative}. Common general-purpose preconditioners such as diagonal (Jacobi) scaling or incomplete LU factorizations~\cite{saad2003iterative} would require recomputation for each topology. Rather than a generic factorization-based preconditioner, we exploit the structure of the contingency problem and use a preconditioner $\matr{P}$ built from a reference system, whose sparsity pattern is shared across all topologies. Explicit forms of the preconditioner $\matr P$ are detailed in Sec.~\ref{sec:algorithm} and \ref{sec:spectral}. The preconditioned operator is $\matr{B}' = \matr{P}^{-1}\matr{B}$, and the preconditioner is effective when $\matr{B}'$ is close to the identity, that is, when the deviation $\matr{A} \coloneqq \eye - \matr{B}'$ is small in norm. The condition number $\kappa(\matr{B}')$ serves as a convenient scalar measure of this closeness, with $\kappa \to 1$ indicating a near-identity operator.

%=============================================================================0

\section{Jacobian decompositions}
\label{sec:jacobian-decompositions}

The key difficulty in parallelizing the NR solve across many topologies on a GPU is that the Jacobian $\matr{J}_t(\vec{x}_t)$ depends on both the topology $t$ and the state $\vec{x}_t$, which vary across scenarios. To enable efficient parallelization, we decompose $\matr{J}_t(\vec{x}_t)$ into a reference part that is shared across all topologies, a low-rank correction that is included with the Woodbury matrix identity and a perturbation that is handled by an iterative solver. The reference part has a fixed sparsity pattern for all $t$, which is the prerequisite for efficient parallelization on a GPU.

We fix a reference topology $r$ and a reference state $\vec{x}_r$, i.e., the base-case $N\!-\!0$ solution, and rewrite the Jacobian for topology $t\in\mathcal{T}$ as
\begin{equation}
    \matr{J}_t(\vec{x}_t) =
    \underbrace{\matr{J}_r(\vec{x}_r)}_{\text{reference}} +\; \matr{R}_t\,\matr{K}^{(\alpha)}_t(\vec{x}_t)\,\matr{R}_t^\top +\; \Delta \matr{J}^{(\alpha)}_r(\vec{x}_t),
    \label{eq:jacobian-decomp-general}
\end{equation}
where $\matr{R}_t \in \mathbb{R}^{(2n+m)\times 4}$ is the sparse topology-change matrix, $\matr{K}^{(\alpha)}_t \in \mathbb{R}^{4\times 4}$ is a small dense matrix, both similar to the definition in Sec.~\ref{sec:background-lowrank}, and $\Delta \matr{J}^{(\alpha)}_r(\vec{x}_t)$ is a perturbation. As described in the following, we can make different choices for the exact definition of $\matr{K}^{(\alpha)}_t$ and $\Delta \matr{J}^{(\alpha)}_r(\vec{x}_t)$.

% \paragraph{Formulation $\alpha=1$ (baseline):}
\textit{Formulation} $\alpha=1$ \textit{(baseline):}
The entire difference from the reference is absorbed into the perturbation:
\begin{equation}
    \matr{K}^{(1)}_t = 0,
    \qquad
    \Delta \matr{J}^{(1)}_r(\vec{x}_t) = \matr{J}_t(\vec{x}_t) - \matr{J}_r(\vec{x}_r).
\end{equation}
No low-rank correction is applied. The perturbation contains both the topology change and the state deviation and thus is generally large.

\textit{Formulation} $\alpha=2$ \textit{(physical method):}
The low-rank term captures the topology change $r\rightarrow t$ at the contingency state $\vec{x}_t$ and the perturbation retains only the state-induced change in the reference topology:
\begin{align}
    \matr{J}_t(\vec{x}_t)
    &= \matr{J}_r(\vec{x}_r)
    + \underbrace{\matr{J}_t(\vec{x}_t) - \matr{J}_r(\vec{x}_t)}_{
        =\,\matr{R}_t\,\matr{K}^{(2)}_t(\vec{x}_t)\,\matr{R}_t^\top}
    + \underbrace{\matr{J}_r(\vec{x}_t) - \matr{J}_r(\vec{x}_r)}_{
        =\,\Delta \matr{J}^{(2)}_r(\vec{x}_t)}.
\end{align}
The perturbation $\Delta \matr{J}^{(2)}_r$ has the same sparsity pattern as $\matr{J}_r$ and is small whenever the contingency state $\vec{x}_t$ is close to the reference state $\vec{x}_r$.

\textit{Formulation} $\alpha=3$ \textit{(physical method with terminal correction):}
This is an extension of $\alpha=2$. The entries of $\Delta \matr{J}^{(2)}_r$ corresponding to the terminal nodes of the failing branch, where the impact of the outage is largest, are moved completely into the low-rank term. This yields a modified $\matr{K}^{(3)}_t$ and a perturbation $\Delta \matr{J}^{(3)}_r$ with smaller entries than $\Delta \matr{J}^{(2)}_r$. In principle this alters the sparsity pattern of the perturbation. We retain the vacated positions as explicit zeros to keep the pattern fixed across all contingencies. Since the matrix enters only through matrix products, the stored zeros carry no cost beyond storage.

\section{Parallel solving algorithm}
\label{sec:algorithm}

Given the decomposition~\eqref{eq:jacobian-decomp-general}, the
NR update $\Delta \vec{x}_t$ is determined by
\begin{equation}
    \left[
    \matr{J}_r(\vec{x}_r)
    + \matr{R}_t \matr{K}^{(\alpha)}_t \matr{R}_t^\top
    + \Delta \matr{J}^{(\alpha)}_r
    \right] \Delta \vec{x}_t = -\vec{f}_t(\vec{x}_t).
\end{equation}
The three summands suggest two natural strategies for organizing the solve. One may first apply the Woodbury matrix identity to invert $\matr{J}_r(\vec x_r) + \matr{R}_t \matr{K}^{(\alpha)}_t (\vec x_t)\matr{R}_t^\top$ exactly and then treat $\Delta \matr{J}^{(\alpha)}_r(\vec x_t)$ as a perturbation handled by an iterative solver. We refer to this as the \textit{Woodbury-first} ordering. Alternatively, one may first handle $\Delta \matr{J}^{(\alpha)}_r(\vec x_t)$ iteratively using $\matr{J}_r(\vec x_r)$ as a preconditioner and then apply the Woodbury correction. We refer to this as the \textit{iterative-first} ordering. The two orderings differ in which part of the operator is inverted exactly and which is left to the iterative solver and hence in the iteration matrix that governs convergence. We derive both orderings below and analyze their spectral properties in Sec.~\ref{sec:spectral}. In each ordering, the reference inverse $\matr{J}_r(\vec{x}_r)^{-1}$ is computed once and reused across all topologies and NR iterations, so that the per-topology factorization of conventional PF is replaced by this shared inversion, a low-rank Woodbury correction, and an iterative solve.

\subsection{Woodbury-first}
\begin{algorithm}[tb]
\caption{Woodbury-first, single NR step using Neumann approximation, all $t \in \mathcal{T}$ in parallel. For the {Bi-CGSTAB} variant replace the Neumann inner loop (steps 5--8) by a preconditioned {Bi-CGSTAB} solve.}
\label{alg:woodbury-first}
\begin{algorithmic}[1]
\Require reference inverse $\matr{J}_r(\vec x_r)^{-1}$ (precomputed once)
\For{$t \in \mathcal{T}$ in parallel}
    \Comment{parallel over topologies}
    \State  compute residual $\vec{f}_t(\vec{x}_t)$, perturbation $\Delta \matr{J}^{(\alpha)}_r(\vec x_t)$
        \Statex \hspace{\algorithmicindent}and low-rank factors $\matr{R}_t$, $\matr{K}^{(\alpha)}_t$
    \State $\matr{S}_t \gets \matr{J}_r^{-1} \matr{R}_t$ \Comment{only once}
    \State $\vec{z}_0 \gets -\vec{f}_t(\vec{x}_t), \quad \Delta\vec{x}_t \gets \vec{0}$
    \For{$k = 0, 1, \ldots, N_k - 1$}
        \Comment{Neumann series}
        \State $\vec{y}_k \gets \matr{P}_t^{-1}\vec{z}_k$ \Comment{via Woodbury}
        \State $\Delta\vec{x}_t \gets \Delta\vec{x}_t + \vec{y}_k$
        \State $\vec{z}_{k+1} \gets -\,\Delta \matr{J}^{(\alpha)}_r(\vec x_t) \, \vec{y}_k$
    \EndFor
\EndFor
\end{algorithmic}
\end{algorithm}

For the Woodbury-first ordering, the reference Jacobian and the low-rank correction are grouped into the matrix that is inverted exactly, $\matr P_t = \matr J_r(\vec{x}_r) + \matr R_t \matr K^{(\alpha)}_t(\vec x_t) \matr R_t^\top$. Left-multiplying the NR update equation by $\matr P_t^{-1}$ gives
\begin{equation}
    \left[ \eye + \matr P_t^{-1} \Delta \matr{J}^{(\alpha)}_r(\vec x_t)
    \right] \Delta \vec{x}_t = -\matr P_t^{-1} \vec{f}_t(\vec{x}_t),
    \label{eq:woodbury-first-precond}
\end{equation}
which is solved with the Neumann series approximation
\begin{equation}
    \Delta \vec x_t = -\left(\sum_{k=0}^{N_k-1}\left(-\matr P_t^{-1}\Delta \matr J_r^{(\alpha)}(\vec x_t)\right)^k\right)\matr P_t^{-1} \vec f_t(\vec x_t).
\end{equation}

Similarly, Eq.~\eqref{eq:woodbury-first-precond} can be viewed as preconditioned system and solved with {Bi-CGSTAB}. In both cases, each application of $\matr P_t^{-1}$ to a vector $\vec{v}$ is computed via the Woodbury matrix identity, 
\begin{multline}
    \matr P_t^{-1} \vec v =
    \matr J_r(\vec x_r)^{-1} \vec v \\
    - \matr S_t \left[ \eye + \matr K^{(\alpha)}_t \matr R_t^\top \matr S_t \right]^{-1} \matr K^{(\alpha)}_t \matr R_t^\top \matr J_r(\vec x_r)^{-1} \vec v,
    \label{eq:woodbury-apply}
\end{multline}
where $\matr{S}_t = \matr{J}_r(\vec{x}_r)^{-1} \matr{R}_t$. The only remaining solve is the $4\times 4$ system in $\eye + \matr K^{(\alpha)}_t \matr R_t^\top \matr S_t$, negligible compared to the other steps.
The complete Woodbury-first algorithm for a single NR step is given in Algorithm~\ref{alg:woodbury-first}.

Steps 5--8 realize the Neumann series, with the inner application of $\matr P_t^{-1}$ in step~6 evaluated through the Woodbury identity~\eqref{eq:woodbury-apply}. The {Bi-CGSTAB} variant instead solves the preconditioned system~\eqref{eq:woodbury-first-precond}, again applying $\matr P_t^{-1}$ via Eq.~\eqref{eq:woodbury-apply}.

\subsection{Iterative-first}

\begin{algorithm}[tb]
\caption{Iterative-first, single NR step using Neumann approximation, all $t \in \mathcal{T}$ in parallel. For the {Bi-CGSTAB} variant replace the Neumann inner loop (steps 5--8) by a preconditioned {Bi-CGSTAB} solve.}
\label{alg:iterative-first}
\begin{algorithmic}[1]
\Require reference inverse $\matr{J}_r^{-1}(\vec x_r)$ (precomputed once)
\For{$t \in \mathcal{T}$ in parallel}
    \Comment{parallel over topologies}
    \State compute residual $\vec{f}_t(\vec{x}_t)$, perturbation $\Delta \matr{J}^{(\alpha)}_r(\vec x_t)$
        \Statex \hspace{\algorithmicindent}and low-rank factors $\matr{R}_t$, $\matr{K}^{(\alpha)}_t$
    \State stack RHS $\matr{Z}_0 \gets \bigl[\, -\vec{f}_t(\vec{x}_t) \;\big|\; \matr{R}_t \,\bigr]$
        \Comment{$5$ columns}
    \State $\matr{G} \gets \matr{0}$
        \Comment{accumulated $\matr{\Xi}_t^{-1}[\,-\vec{f}_t \mid \matr{R}_t\,]$}
    \For{$k = 0, 1, \ldots, N_k - 1$}
        \Comment{Neumann series}
        \State $\matr{Y}_k \gets \matr{J}_r^{-1} \matr{Z}_k$
            \Comment{dense GEMM}
        \State $\matr{G} \gets \matr{G} + \matr{Y}_k$
        \State $\matr{Z}_{k+1} \gets -\,\Delta \matr{J}^{(\alpha)}_r(\vec x_t) \, \matr{Y}_k$
            \Comment{sparse SpMM}
    \EndFor
    \State $\vec{g}_t \gets \matr{G}_{[:,1]}, \quad
        \matr{W}_t \gets \matr{G}_{[:,2:5]}$
        \Comment{split columns}
    \State solve $4\times4$:
        $\bigl[\eye + \matr K^{(\alpha)}_t\matr{R}_t^\top \matr{W}_t\bigr]\,\vec{z}_t
        = \matr K^{(\alpha)}_t\matr{R}_t^\top \vec{g}_t$
    \State $\Delta\vec{x}_t \gets \vec{g}_t -\matr{W}_t \vec{z}_t$
        \Comment{Woodbury combine}
\EndFor
\end{algorithmic}
\end{algorithm}

\begin{figure*}[tb]
    \centering
    \includegraphics[width=\linewidth]{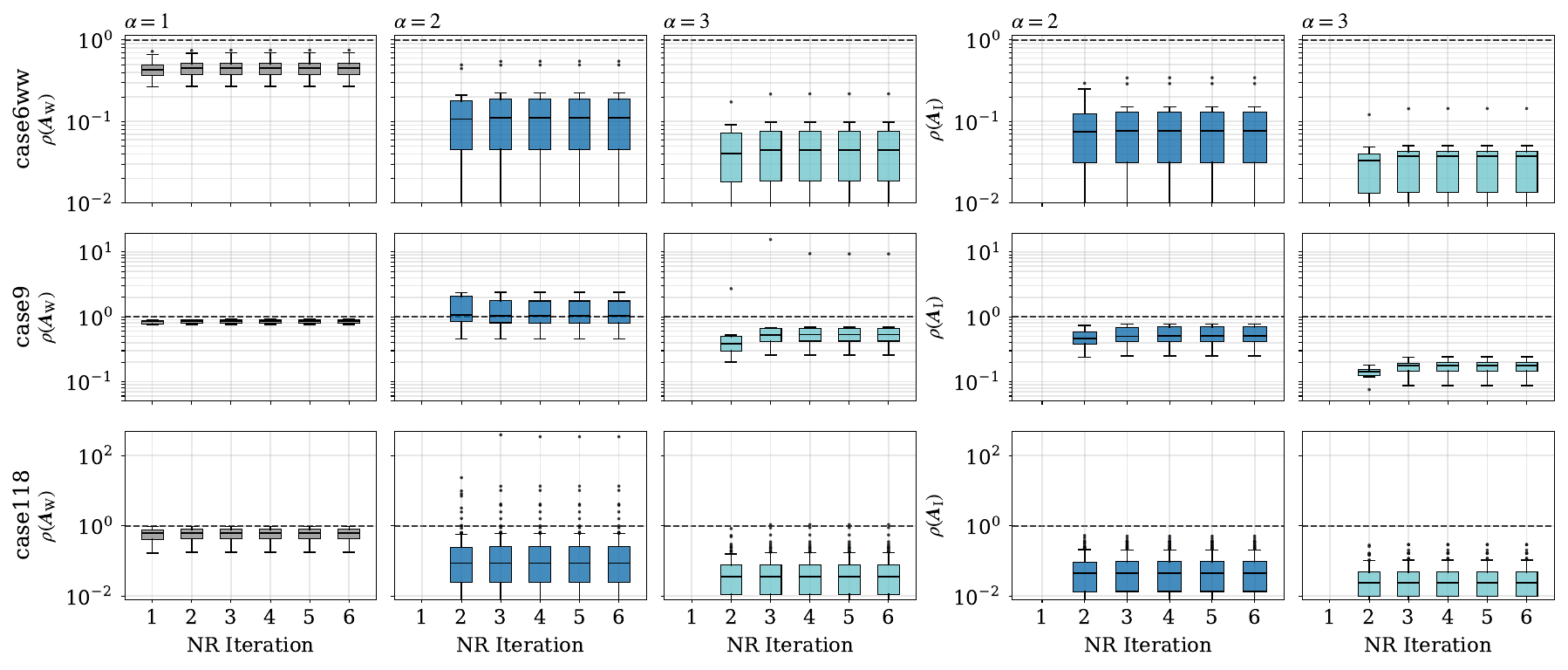}
    \caption{
    Spectral radius $\rho$ of the iteration matrix, $\matr A_\mathrm{W}= -\matr P_t^{-1}\, \Delta \matr J^{(\alpha)}_r $ and $\matr A_\mathrm{I}= -\matr J_r(\vec x_r)^{-1}\, \Delta \matr J^{(\alpha)}_r$, for decompositions $\alpha = 1,2,3$ shown for all $N\!-\!1$ contingencies and Newton--Raphson iterations for test cases case6ww, case9 and case118 (rows). Left three columns: Woodbury-first ordering. Right two columns: iterative-first ordering. Since $\alpha = 1$ yields identical results for both orderings, it is shown only in the left block. The dashed line marks $\rho = 1$, above which the Neumann series diverges.
    }
    \label{fig:all_rhos}
\end{figure*}
\begin{figure*}[tb]
    \centering
    \includegraphics[width=\linewidth]{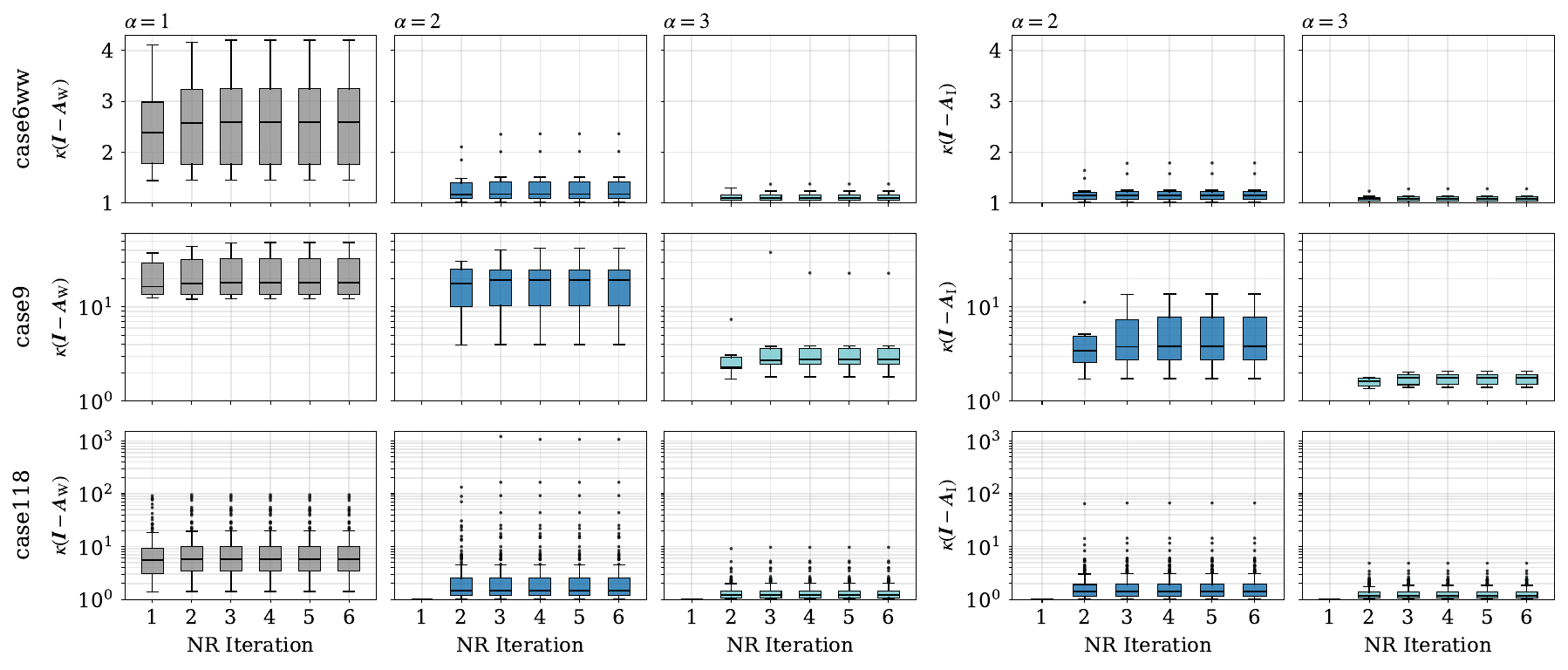}
    \caption{Condition number $\kappa$ of the preconditioned system $\eye - \matr A$, with $\matr A_\mathrm{W}$ and $\matr A_\mathrm{I}$ as defined in \eqref{eq:iteration_matrix_woodbury_first} and \eqref{eq:iteration_matrix_iterative_first}, for the same combinations and panel ordering as Fig.~\ref{fig:all_rhos}.}
    \label{fig:all_conds}
\end{figure*}

Now, instead we only use the reference Jacobian $\matr J_r(\vec x_r)$ as a preconditioner and group the reference Jacobian and the perturbation together,
\begin{equation}
    \matr \Xi_t = \matr{J}_r(\vec{x}_r)
    + \Delta \matr{J}^{(\alpha)}_r(\vec x_t),
\end{equation}
so that the update equation reads
\begin{equation}
    \left[ \matr \Xi_t
    + \matr{R}_t \matr{K}^{(\alpha)}_t(\vec x_t) \matr{R}_t^\top \right]
    \Delta \vec{x}_t = -\vec{f}_t(\vec{x}_t).
\end{equation}
The Woodbury identity is now applied using the full inverse of $\matr \Xi_t$,
\begin{equation}
\begin{split}
    \Delta &\vec{x}_t = -\matr \Xi_t^{-1} \vec f_t(\vec x_t)+ \matr \Xi_t^{-1} \matr R_t \\
    &\times\left[ \eye + \matr K^{(\alpha)}_t\matr R_t^\top \matr \Xi_t^{-1} \matr R_t \right]^{-1} \matr K^{(\alpha)}_t\matr R_t^\top \matr \Xi_t^{-1} \vec f_t(\vec x_t).
\end{split}
\label{eq:woodbury-iterative-first}
\end{equation}
In practice, the terms $\matr \Xi_t^{-1} \vec f_t(\vec x_t)$ and $\matr \Xi_t^{-1} \matr R_t$ are not computed separately but stacked column-wise into a single right-hand side $\matr Z_t = [\,-\vec f_t(\vec x_t) \mid \matr R_t\,]$ and solved together by a single batched Neumann iteration,
\begin{align*}
    \matr \Xi_t^{-1} \matr Z_t &= \left(\matr J_r(\vec x_r) + \Delta \matr J^{(\alpha)}_r(\vec x_t)\right)^{-1} \matr Z_t \\
    &= \sum_{k=0}^{N_k - 1} \left( -(\matr J_r(\vec{x}_r))^{-1} \Delta \matr{J}^{(\alpha)}_r(\vec x_t) \right)^k \left(\matr J_r(\vec{x}_r)\right)^{-1} \matr Z_t,
\end{align*}
so that each Neumann term reduces to a dense product with the shared inverse $\matr J_r(\vec x_r)^{-1}$ and a sparse product with $\Delta \matr J^{(\alpha)}_r(\vec x_t)$, both applied to all columns of $\matr Z_t$ at once. This batching, together with the parallelization over topologies $t \in \mathcal{T}$, is what maps the solve efficiently onto the GPU.

Note that with a hot-start $\vec x_t^{(1)} = \vec x_r$, the perturbation $\Delta \matr J_r^{(\alpha)}(\vec x_t)$ vanishes at the first NR iteration for $\alpha=2,3$. Thus, for both orderings the first NR step reduces to a single exact Woodbury correction with no inner iterations.

The complete iterative-first algorithm is given in Algorithm~\ref{alg:iterative-first}. Steps 5--8 realize the batched Neumann iteration described above and steps 9--11 perform the rank-four Woodbury step. The {Bi-CGSTAB} variant replaces the inner loop (steps 5--8) by a batched preconditioned {Bi-CGSTAB} solve with the preconditioner $\matr{J}_r(\vec x_r)$.

\section{Spectral radius and condition number}
\label{sec:spectral}
Both orderings reduce the Newton step to a preconditioned system of the same structure but with a different preconditioner $\matr{P}$, which we keep generic in the following derivation and specialize to each ordering afterward. Writing the iteratively solved part of the system as $\matr P\,\Delta\vec{x}_t + \Delta \matr J^{(\alpha)}_r(\vec x_t)\,\Delta\vec x_t = -\vec f_t(\vec x_t)$ and left-multiplying by $\matr P^{-1}$ gives
\begin{equation}
(\eye - \matr A)\,\Delta\vec x_t = -\matr P^{-1}\vec f_t,
\quad
\matr A \coloneqq -\matr P^{-1}\Delta \matr J^{(\alpha)}_r(\vec x_t),
\label{eq:iteration-matrix}
\end{equation}
where $\matr{A}$ is the iteration matrix whose spectral properties govern convergence. The two orderings correspond to two choices of $\matr{P}$.
In the Woodbury-first ordering the preconditioner $\matr P_t = \matr J_r(\vec{x}_r) + \matr R_t \matr K^{(\alpha)}_t(\vec x_t) \matr R_t^\top$ is applied exactly via the Woodbury identity~\eqref{eq:woodbury}, giving the iteration matrix
\begin{equation}
    \matr A_{\mathrm{W}} = -\matr P_t^{-1}\, \Delta \matr J^{(\alpha)}_r(\vec x_t),
    \label{eq:iteration_matrix_woodbury_first}
\end{equation}
so that $\eye - \matr A_{\mathrm{W}} = \eye + \matr P_t^{-1} \Delta \matr J^{(\alpha)}_r(\vec x_t)$ recovers the preconditioned system of Eq.~\eqref{eq:woodbury-first-precond}.
In the iterative-first ordering the reference Jacobian $\matr J_r(\vec x_r)$ alone serves as preconditioner, giving the iteration matrix
\begin{equation}
    \matr A_{\mathrm{I}} = -\matr J_r(\vec x_r)^{-1}\, \Delta \matr J^{(\alpha)}_r(\vec x_t),
    \label{eq:iteration_matrix_iterative_first}
\end{equation}
and leaving the low-rank term to a subsequent Woodbury correction.

The efficiency of both iterative solvers is governed by the spectral properties of $\matr{A}$: the Neumann series requires $\rho(\matr{A}) < 1$, while for the preconditioned \mbox{Bi-CGSTAB} the preconditioner quality is studied through the condition number $\kappa(\eye-\matr A)$ of the preconditioned system, keeping in mind that this gives only a coarse estimate of the convergence.

We therefore study $\rho(\matr A)$ and $\kappa(\eye-\matr A)$ for all three decompositions $\alpha = 1,2,3$ combined with both orderings, on the standard benchmark cases case6ww, case9 and case118, imported via pandapower~\cite{thurner2018pandapower, zimmermann2011matpower}. 
This spectral analysis is restricted to small systems because it requires computing eigenvalues and singular values of the iteration matrix for every contingency and NR iteration, whose costs scale as $\mathcal{O}(N_m^3)$, which becomes intractable for large grids.
To characterize the iteration matrix independently of any particular inner iterative solver, the update step itself is computed with an exact linear solve of the full contingency Jacobian $\matr J_t (\vec x_t)$, then the spectral quantities are evaluated at each true NR iterate. The spectral radii across all $N\!-\!1$ contingencies and NR iterations are shown in Fig.~\ref{fig:all_rhos} and the corresponding condition numbers in Fig.~\ref{fig:all_conds}. For $\alpha = 1$ we only show one ordering since both orderings coincide. For $\alpha=2,3$ the perturbation $\Delta \matr J_r^{(\alpha)}(\vec x_t)$ vanishes at the first NR iteration (Sec.~\ref{sec:algorithm}) due to the hot start with $\vec x_r$, so no box appears at the first iteration. 

The spectral radii for $\alpha = 1$ cluster just below unity across all NR iterations, confirming that the baseline decomposition is unsuitable for a fast converging Neumann solve. Both $\alpha=2$ and $\alpha=3$ reduce $\rho$ substantially, with $\alpha = 3$ achieving the lowest values across both orderings. The iterative-first ordering yields systematically lower spectral radii than the Woodbury-first ordering for the same $\alpha$. A small number of outlier contingencies with $\rho > 1$ persist for the Woodbury-first ordering, particularly in case118, corresponding to cases where the state deviation introduced due to the topology change is large and better handled after the iterative inner solver.

The condition numbers exhibit the same qualitative behavior as the spectral radii: $\alpha = 1$ yields poorly conditioned systems, while $\alpha=2$ and $\alpha=3$ lower the bulk of the condition-number distribution, with $\alpha=3$ giving the strongest reduction, while, again, the iterative-first ordering outperforms Woodbury-first for a given $\alpha$. The lowest condition numbers are achieved by $\alpha = 3$ with the iterative-first ordering.
 
This agreement across the two metrics is central to the results of this analysis. A small $\rho$ ensures rapid convergence of the Neumann series, while a small $\kappa$ reflects a well-conditioned system for the \mbox{Bi-CGSTAB} iteration. The fact that both are minimized by the same configuration means a single choice optimizes both iterative solvers and thus constitutes the best preconditioning strategy. We therefore adopt $\alpha = 3$ with the iterative-first ordering for all subsequent numerical experiments.

\begin{figure}[tb]
    \centering
    \includegraphics[width=\linewidth]{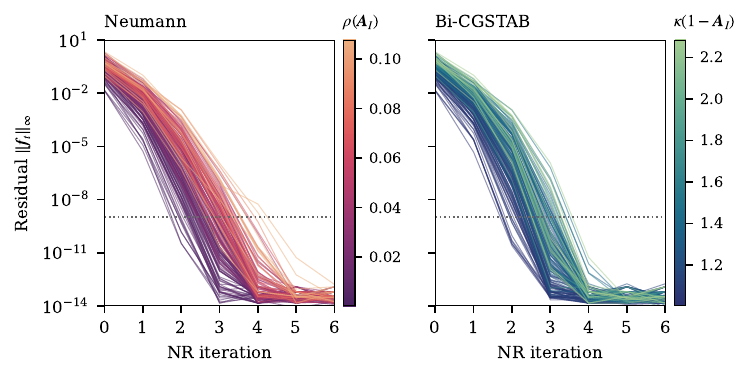}
    \caption{Residual $\|\vec f_t\|_\infty$ versus NR iteration for the Neumann (left) and \mbox{Bi-CGSTAB} (right) inner solves on case118, at inner-iteration count $N_k = 5$. Each line is one $N\!-\!1$ contingency, colored by the spectral radius $\rho(\matr A_\mathrm{I})$ (left) and condition number $\kappa(\eye - \matr A_\mathrm{I})$ (right) of its iteration matrix. The dashed line marks the convergence threshold $\|\vec f_t\|_\infty < 10^{-9}$. Residuals plateau near machine precision ($\sim 10^{-14}$).}    \label{fig:achieved_residual_rho}
\end{figure}
To confirm that these spectral quantities translate into real achieved convergence, Fig.~\ref{fig:achieved_residual_rho} shows the residual trajectories of the actual Neumann and \mbox{Bi-CGSTAB} inner solves on case118, at the same inner-iteration count $N_k=5$ adopted for the performance evaluation in Sec.~\ref{sec:performance}. The spectral quantities are evaluated along the exact NR trajectory (cf.~Fig.~\ref{fig:all_rhos} and Fig.~\ref{fig:all_conds}), whereas the residuals are those of the actual inner solver. The trajectories are colored by $\rho(\matr A_I)$ (Neumann) and $\kappa(\eye - \matr A_I)$ (\mbox{Bi-CGSTAB}). Both patterns confirm the spectral analysis: the contingencies with the largest spectral radii (Neumann) and the largest condition numbers (\mbox{Bi-CGSTAB}) converge visibly slower, while the remainder reach near machine precision (residual $\sim 10^{-14}$) within fewer iterations. The rate separation is most pronounced for the Neumann series, whose geometric convergence factor $\rho$ directly sets the per-iteration reduction. For \mbox{Bi-CGSTAB}, where no single scalar fully characterizes the convergence (see Sec.~\ref{sec:background-tools}), the separation by condition number is less sharp. Nonetheless, a trend remains visible, confirming that $\kappa$ provides a useful, though approximate, indicator of convergence behavior.

\section{Computational performance}
\label{sec:performance}

Having identified $\alpha = 3$ with the iterative-first ordering as the most effective configuration in Sec.~\ref{sec:spectral}, we now evaluate its computational performance. 
The algorithm is implemented in Julia~\cite{bezanson2017julia} using the CUDA.jl package~\cite{besard2018juliagpu} and benchmarked on an NVIDIA H100 GPU. As CPU baselines we use LightSim2Grid on a single i7-13700H core~\cite{lightsim2grid}, and 80-core implementations of LightSim2Grid and pandapower~\cite{thurner2018pandapower} on an AMD EPYC 9V84, with numbers taken from~\cite{faber2025thesis}. All results are reported for the case2869pegase test case~\cite{josz2016acpowerflowdata, fliscounakis2013contingency} with 2869 buses representing a realistic transmission-scale $N\!-\!1$ study. Consistent with using the reference Jacobian as preconditioner, we hot-start the NR iteration from the reference state $\vec x_r$, the $N\!-\!0$ solution. The implementation used to produce the results in this paper is available at~\cite{code_low_rank_with_inverse}.

\subsection{Per-iteration runtime breakdown}

\begin{figure}[tb]
    \centering
    \includegraphics[width=\linewidth]{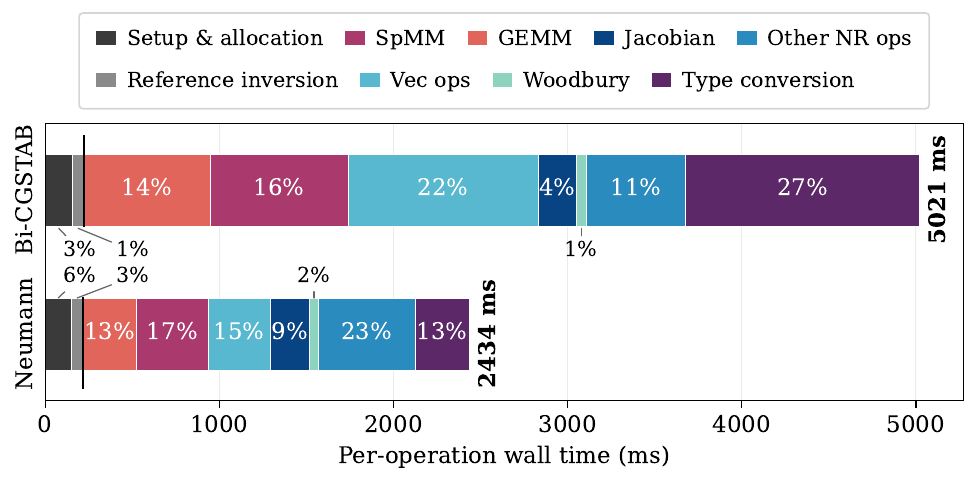}
    \caption{Whole solver wall-time composition on case2869pegase (NVIDIA H100, $(N_n,N_k)=(5,5)$), for the Neumann and \mbox{Bi-CGSTAB} inner solvers. Costs are grouped and split at the divider into one-time work (left: data loading, allocation and the reference Jacobian inversion) and per-iteration work (right). Percentages give each group's share of the total wall time, the bold number is the total wall time. Because the breakdown is measured with per-operation synchronization, the segments sum to the bar total, which exceeds the overlapped production wall time reported in Sec.~\ref{sec:performance-comparison}.}
    \label{fig:bar_chart_single_NR}
\end{figure}

To identify where the runtime is spent and to guide further optimization, Fig.~\ref{fig:bar_chart_single_NR} decomposes the total wall time of a complete $N\!-\!1$ solve into grouped components, separated into one-time and per-iteration costs. Two points should be noted. First, the breakdown is a profiling run in which each operation is measured under synchronization. The segments therefore sum to a total, but that total exceeds the overlapped wall time of the production solve (Sec.~\ref{sec:performance-comparison}), where the GPU scheduler overlaps independent operations. Second, the reported time is the total across all contingencies, not the per-topology time used in the CPU comparison. The breakdown is thus meant to show the \emph{relative} distribution of work, not absolute performance.

The one-time costs are dominated by setup and allocation. The $\mathcal{O}(N_m^3)$ inversion of the reference Jacobian accounts for only a small share, as $\matr{J}_r(\vec{x}_r)^{-1}$ is computed once and reused across every contingency and NR iteration. We apply it through a Tensor-Core  dense matrix--matrix product (GEMM) in the reduced-precision TF32 format (TensorFloat-32), roughly doubling the throughput of the dominant dense product, while the outer Newton iteration and the low-rank correction remain in double precision (FP64). The preconditioner is stored once in single precision (FP32) and at each application its operands are cast down from FP64 and the result cast back up, a type conversion visible in the breakdown. The TF32 GEMM nonetheless outperforms the equivalent FP32 or FP64 product, so this overhead is recovered.

Apart from the type conversions, the per-iteration cost is dominated by the linear algebra: the GEMM applying the reference inverse and the sparse matrix--matrix product (SpMM) applying the perturbation together form the largest share for both solvers. \mbox{Bi-CGSTAB} additionally spends a comparable amount in its vector recurrences (dot products and updates), which the Neumann series does not require. The Jacobian assembly is a small to moderate per-iteration contributor (4.34\% and 9.37\%, respectively), and the residual and edge-term evaluations are smaller still, summarized under the label \textit{Other NR steps}, together with the calculation of matrices $\matr K_t^{(3)}(\vec x_t)$ and $\Delta\matr J_r^{(3)}(\vec x_t)$, the state vector update itself, among other small contributors.
In contrast to the whole solver workload, the rank-four Woodbury correction is negligible, contributing only $2.06\%$ of the runtime for Neumann and $1.14\%$ for \mbox{Bi-CGSTAB}.

\subsection{Iteration count trade-off}
\begin{figure}[tb]
    \centering
    \includegraphics[width=\linewidth]{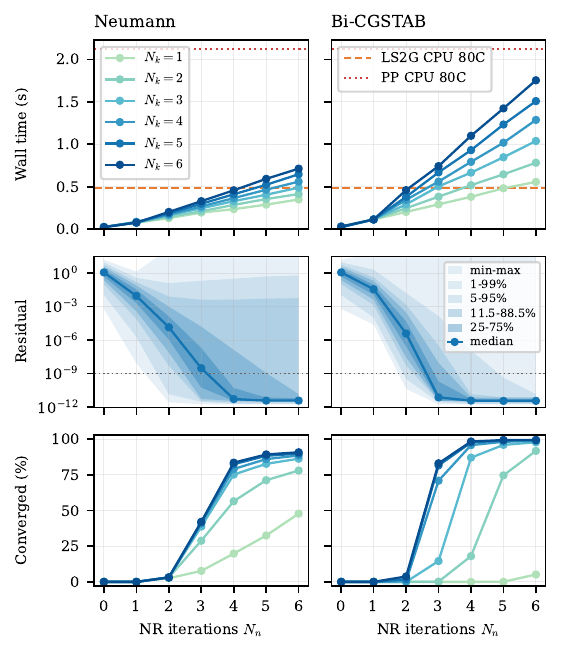}
    \caption{Wall time (top row), residual distribution for $N_k=5$ (middle row) and percentage of converged topologies (bottom row) for Neumann (left column) and {Bi-CGSTAB} (right column) as inner solvers on case2869pegase, for varying numbers of NR iterations $N_n$ (horizontal axis). In the top and bottom rows, colors denote the inner-solver iteration count $N_k$ and in the middle row, the shaded bands show percentile ranges of the final residual (see legend), at the fixed $N_k=5$ with the dotted line marking the convergence threshold on the residual. In the top row, dashed (dotted) horizontal lines mark the LightSim2Grid (pandapower) 80-core CPU baselines.
    }
    \label{fig:sweep_neumann}
\end{figure}
Before comparing against CPU baselines, we determine the operating point $(N_n, N_k)$, i.e., the number of outer NR iterations $N_n$ and inner solver (Neumann or \mbox{Bi-CGSTAB}) iterations $N_k$, that achieves acceptable convergence. Convergence for a contingency case is defined as a maximum residual $\|\vec f_t(\vec x_t)\| < 10^{-9}$. Figure~\ref{fig:sweep_neumann} shows the wall time (top row), the distribution of the residual using median, quartiles and outer percentiles (middle row) and the percentage of converged cases (bottom row). Three observations follow.

First, for a sufficiently accurate inner solve (large $N_k$, best seen for \mbox{Bi-CGSTAB}), four outer NR iterations suffice to converge the large majority of contingency cases. This confirms the $N\!-\!1$ states are close enough to the reference state for the outer loop to converge rapidly. To ensure convergence with some margin, we set $(N_n, N_k) = (5, 5)$ in the following.

Second, at $(N_n, N_k) = (5, 5)$ the Neumann solver reaches wall times competitive with the LightSim2Grid 80-core CPU baseline, but does not converge for all cases: it reaches roughly $90\,\%$ coverage, as a fraction of contingencies have iteration-matrix spectral radii exceeding unity, for which the Neumann series diverges. This makes the Neumann solver suitable for fast screening, where incomplete coverage is acceptable, but not for exact $N\!-\!1$ assessment.

Third, the \mbox{Bi-CGSTAB} solver reaches about $99\,\%$ convergence across the $N\!-\!1$ cases, matching the coverage of the CPU baseline, but at wall times exceeding the 80-core LightSim2Grid baseline.
This is due to \mbox{Bi-CGSTAB}'s higher numerical cost per iteration, from the additional matrix products and vector operations within each step. Across all configurations, both solvers stay below the pandapower 80-core benchmark.

\subsection{Comparison to CPU baselines}
\label{sec:performance-comparison}

Table~\ref{tab:per_topology_times} reports the per-topology time on case2869pegase, introduced above, together with two further test cases, case118 and case1354pegase~\cite{josz2016acpowerflowdata, fliscounakis2013contingency}, at the fixed operating point ${(N_n, N_k) = (5, 5)}$. We report per-topology rather than total wall time so that grids of different size can be compared on a common basis. Additionally, the GPU solvers and the 80-core CPU baselines are evaluated on the exact same set of contingencies. Both measurements cover the same stages of the computation: the timed region is the solver itself, the repeated linear solves, and excludes the one-time setup of reading the grid data, allocation and, for the GPU, host--device transfer. This overhead differs between implementations and can be reduced substantially in a production setting~\cite{faber2025thesis}. The dense reference inverse's memory footprint is modest: at single precision it is about {$104$\,MB} for case2869pegase, and even for the much larger case9241pegase it remains around {$1$\,GB}. The constraint on how many contingencies can be solved in parallel is therefore the batched working set, not the inverse.

\begin{table}[t]
  \centering
  \caption{Per-topology wall time (\textnormal{ms}) for $N\!-\!1$ security assessment at $(N_n, N_k) = (5, 5)$ on three test grids. The GPU Neumann and \mbox{Bi-CGSTAB} variants (NVIDIA H100) are compared against LightSim2Grid (LS2G) and pandapower on 80 CPU cores and LS2G on a single core. Core counts are indicated as (80c)/(1c). The fastest result per case is highlighted in bold. CPU baselines are taken from~\cite{faber2025thesis} (80-core) and~\cite{lightsim2grid} (single-core).}
  \label{tab:per_topology_times}
\begin{tabular}{lrrr}
    \toprule
     & \textbf{case118} & \textbf{case1354pegase} & \textbf{case2869pegase} \\
    \# $N\!-\!1$ cases & 177 & 1430 & 3804 \\
    \midrule
    GPU Neumann   & \textbf{0.097} & \textbf{0.092} & 0.155 \\
    GPU Bi-CGSTAB  & 0.162 & 0.185 & 0.342 \\
    LS2G (80c)     & --    & 0.203 & \textbf{0.126} \\
    pandapower (80c) & 1.53 & 0.434 & 0.557 \\
    LS2G (1c)      & 0.047 & 1.06  & 2.35 \\
    \bottomrule
  \end{tabular}
\end{table}

The GPU Neumann solver undercuts the LightSim2Grid 80-core per-topology time on case1354pegase (no 80-core LightSim2Grid value is available for case118), and stays close to it on case2869pegase. The \mbox{Bi-CGSTAB} solver remains above the LightSim2Grid 80-core baseline across all cases but consistently undercuts the pandapower 80-core benchmark. The two baselines differ in intent: LightSim2Grid is a performance-optimized C++ PF backend and thus the stronger reference, whereas pandapower is a broader, general-purpose Python tool not specialized for high-throughput contingency analysis.

We also include a single-core baseline. It performs best only on the small case118 grid, where parallelization overhead does not yet pay off. On the two larger grids it is clearly outpaced by both the GPU and the 80-core CPU solvers.

\section{Conclusion}
\label{sec:conclusion}

We have presented a GPU-parallel Newton--Raphson solver for $N\!-\!1$ contingency analysis that separates the power-flow state from the network topology, allowing a single reference factorization to be reused across all contingencies. The reference Jacobian serves as an effective preconditioner for the iterative inner solve, while the topology change is absorbed by a low-rank Woodbury correction, removing the need to refactorize the Jacobian for every contingency.
A spectral analysis confirms the preconditioner’s effectiveness; both the Neumann series and Bi-CGSTAB converge in few iterations, with Neumann faster and Bi-CGSTAB handling the outlier cases where the series diverges.

GPU parallelization of batched contingency analysis is effective but does not render optimized CPU methods obsolete: a single GPU proves comparable to 80 highly optimized CPU cores. The appropriate framing is thus one of complementary hardware rather than order-of-magnitude GPU superiority.

Several directions follow. The runtime breakdown points to the precision conversion and the Bi-CGSTAB vector recurrences as the main optimization targets. Second, fast contingency screening, a cheaper, low-accuracy pass to pre-filter the \textit{a priori} unknown set of critical contingencies, could further reduce the overall cost. Finally, the method extends naturally to further contingency types, such as generator and busbar outages. The latter remove an entire node and thus yield higher-rank corrections.

\bibliographystyle{IEEEtran}
\bibliography{references}

\end{document}